\documentclass[twocolumn]{aastex631}
\usepackage{hyperref}
\usepackage{color}
\usepackage{amssymb}
\usepackage{amsmath}
\usepackage[ruled,linesnumbered]{algorithm2e}
\usepackage{footmisc}
\usepackage{ulem}

\received{January 1, 0000}
\revised{January 1, 0000}
\accepted{January 1, 0000}
\submitjournal{ApJ}

\shorttitle{SNe Ia Ni}
\shortauthors{Chen et al.}
\begin{document}

\title{Artificial Intelligence Assisted Inversion (AIAI): Quantifying the Spectral Features of $^{56}$Ni of Type Ia Supernovae}
\correspondingauthor{Lifan Wang}
\email{lifan@tamu.edu}

\author{Xingzhuo Chen}
\affiliation{George P. and Cynthia Woods Mitchell Institute for Fundamental Physics \& Astronomy, \\
Texas A. \& M. University, Department of Physics and Astronomy, 4242 TAMU, College Station, TX 77843, USA}

\author{Lifan Wang}
\affiliation{George P. and Cynthia Woods Mitchell Institute for Fundamental Physics \& Astronomy, \\
Texas A. \& M. University, Department of Physics and Astronomy, 4242 TAMU, College Station, TX 77843, USA}

\author{Lei Hu}
\affiliation{Purple Mountain Observatory, Nanjing 210008, People's Republic of China}

\author{Peter J. Brown}
\affiliation{George P. and Cynthia Woods Mitchell Institute for Fundamental Physics \& Astronomy, \\
Texas A. \& M. University, Department of Physics and Astronomy, 4242 TAMU, College Station, TX 77843, USA}

\begin{abstract}
    Following our previous study of Artificial Intelligence Assisted Inversion (AIAI) of supernova analyses \citep{Xingzhuo2020AIAI}, we train a set of deep neural networks based on the one-dimensional radiative transfer code TARDIS \citep{tardis} to simulate the optical spectra of Type Ia supernovae (SNe~Ia) between 10 and 40 days after the explosion. The neural networks are applied to derive the mass of $^{56}$Ni in velocity ranges well above the photosphere for a sample of 153 well-observed SNe~Ia. Many SNe have multi-epoch observations for which the decay of the radioactive $^{56}$Ni can be tested quantitatively. 
    The $^{56}$Ni mass derived from AIAI using the observed spectra as input for the sample is found to agree with the theoretical $^{56}$Ni decay rate. The AIAI reveals a spectral signature near 3890 \AA\, which can be identified as being produced by multiple Ni II lines between $3950$ and $4100\ \AA$. 
    The mass deduced from AIAI is correlated to the  light-curve shapes of SNe~Ia, with the SNe~Ia with broader light curves showing larger $^{56}$Ni mass in the envelope.
    AIAI enables spectral data of SNe to be quantitatively analyzed under theoretical frameworks based on well-defined physical assumptions. 
\end{abstract}

\section{Introduction}

Type Ia supernovae (SNe Ia) are used for cosmological distance measurements based on empirical relations between their light curve shapes and luminosities \citep{phillips1993dm15relation,GuySALT,lifan2003LightCurveCalib,rubin2013DistMeasure,YangJiaWen2022ApJ...938...83Y,LaurenAldoroty}. 
These relations are consistent with theoretical models of the thermonuclear explosion of a white dwarf close to the Chandrasekhar mass limit (1.4 $M_\odot$). 
Although several three-dimensional SNe~Ia explosion simulations could provide comparable photometric and spectroscopic features compared to the observed SNe~Ia light curves and spectral sequences \citep[e.g.,][]{Bulla2016doubledet,Wilk2017cmfNew,townsley2019sedonaDoubleDet}, the first principle hydrodynamics, nucleosynthesis, and radiative transfer simulation results of SNe~Ia are extremely expensive to calculate and still far from the level of precision that matches those of observations.

Hydrodynamic and nucleosynthesis processes are only significant in the first $\sim$ 100 seconds\citep{Ropke2005Homologous}. 
Thereafter the supernova (SN) ejecta expand homologously and the observed optical spectra and light curves are governed by radiative transfer and the thermal state of the ejecta. 
Modeling the radiative transfer process is essential in the estimates of the density profiles and elemental abundances of the SN ejecta so as to put constraints on the SN explosion mechanisms. 
Several programs have been developed for the calculations of synthetic spectra. 
The simplest is the SYNOW code  \citep{Parrent2010SYNOW,Thomas2011SYNAPPS} which directly uses line opacity data to generate synthetic spectra  and has been widely used in spectral line identifications. 
Hydra \citep{Hoeflich1996Hydra}, PHOENIX \citep{Haushildt2006PHOENIX} and CMFGEN \citep{Hillier1998CMFGEN} are more advanced simulation programs that allow for more accurate modeling through finite element approach. 
Programs using the Monte-Carlo method have also been developed for spectral simulation (e.g., SEDONA \citep{KasenSedona2006ApJ...651..366K}, ARTIS \citep{Kromer2009Artis}) and have been used for spectropolarimetry calculations \citep{Hoel=flich1991A&A...246..481H,Kasen:2003ApJ...593..788K,Bulla2015Polar,Livneh2022Polar}. 
In spectropolarimetry modeling, 3-D effects, non-local thermodynamic equilibrium (NLTE), and time-dependent radiative transfer effects are important.  
Among these codes, TARDIS \citep{tardis} is a one-dimensional, time-independent Monte-Carlo radiative transfer program with approximate NLTE treatment of atomic processes. 



Model-based analyses enable "abundance tomography" of the SN ejecta \citep[e.g.,][]{Hoeflich1996Hydra,stehle2005abuntomo}. 
The Artificial Intelligence Assisted Inversion (AIAI) of SN~Ia spectra \citep{Xingzhuo2020AIAI} trains a set of neural networks (NNs) based on the simulated spectra using the radiative transfer program TARDIS \citep{tardis} to derive the elemental abundances of several well-observed SNe Ia at around 20 days after the explosion. 
\cite{kerenzdorf2020dalek} used a NN to accelerate the calculation of the radiative transfer program, and suggest that the method can be used for SN ejecta structure estimate based on a nested sampling algorithm. 

Among the 23 elements (from number 6 element C, to number 28 element Ni) that play significant roles in the nucleosynthesis process of the SNe~Ia, $^{56}$Ni stands out for two reasons. 
Firstly, the decay chain of $^{56}$Ni is the major energy source that powers the radiative process of SN~Ia.  
The observed light curves can be used to determine the total $^{56}$Ni masses of SNe~Ia \citep[e.g,][]{arnett1982rule,Woosley2007LC,khatami2019arnett}. 
Secondly, the spatial distribution of $^{56}$Ni in SN ejecta is directly related to the SN explosion mechanism  \citep[e.g.,][]{Piro2010Shock,Piro2016LCDiver}). 
A problem is that unlike the Si II $6355\mathrm{\AA}$ and Ca II $8542\mathrm{\AA}$ lines, which are strong lines that are clearly distinguishable from other lines, the spectral lines of $^{56}$Ni are blended and cannot be easily measured from the observed SN Ia spectra around the maximum. 
\cite{childress2015CoNebula} use Co III  $\lambda 5893$ emission line in the nebular spectra of SNe~Ia to estimate the mass before the radioactive decay of $^{56}$Ni. 
The AIAI approach has the distinct advantage that the line measurements do not rely strongly on any single spectral features. With a large enough training set, it can derive optimal estimates of the physical parameters with the entire observed spectra contributing to the reliability of all the parameters. 




In this paper, we present a set of NNs and apply them to a sample of well-observed SNe~Ia.
The NNs are trained on the simulated spectra using TARDIS. 
Using the NN predicted SN ejecta structure, we identify a correlation between the spectral feature around $3950 \mathrm{\AA}$ of SNe~Ia spectra around B-band maximum time and the $^{56}$Ni mass in the SN ejecta. 
The paper is structured as follows: 
Section \ref{sec:method} introduces the training dataset, the NN architecture, and the spectral sequence fitting pipeline. 
Section \ref{sec:result} shows the results of the spectral line measurement and the Ni mass measurements. 
Section \ref{sec:conclusion} gives the conclusions. 
The code for this research is available at \href{https://github.com/GeronimoChen/DLTD}{https://github.com/GeronimoChen/DLTD}. 

\section{Method}\label{sec:method}

\subsection{TARDIS spectral calculation}

In TARDIS calculation, SN ejecta is assumed to be spherically symmetric and expands homogeneously after the explosion. 
To avoid the computationally demanding time-dependent photon diffusion calculation, a simple approximation of the photosphere with a black-body emission spectrum at the inner boundary of SN ejecta is adopted.  
Given an SN ejecta structure, a target luminosity, a time after the explosion, and an inner boundary velocity, TARDIS can calculate a spectrum in $\sim$ 1 CPU hour. 

The model spectra are calculated using the code TARDIS (version 3.0-dev3448). 
To set up the models, we use dilute local thermodynamic equilibrium ({\tt\string dilute-lte})\cite{Lucy1999downbranch} approximation to calculate the atomic level population, nebular approximation ({\tt\string nebular})\cite{mazzali1993nebular} to calculate the ionization fraction, {\tt\string macroatom}\cite{Lucy2002Macroatom,Lucy2003Macroatom2} to calculate the atomic level transition, {\tt\string detailed} to calculate the black-body radiative rates. 
TARDIS uses an iterative algorithm \citep{Lucy2003Macroatom2} to calculate a self-consistent SN temperature profile. 
As in our previous work on AIAI of SNe~Ia spectra \cite{Xingzhuo2020AIAI} (Hereafter XLL20), the step of temperature iterations is set to 15, with the damping parameters being: ({\tt\string type:damped. damping\_constant:1 threshold:0.05. fraction:0.8. hold\_iterations:3. t\_inner\_damping\_constant:1}). 
The temperature calculation setup in our research does not return a converged temperature profile, but the final temperature profile can be modified by changing the target luminosity and the location of the inner boundary. 
We adopt this configuration for its flexibility in the temperature calculation  which is necessary when the observed and the simulated spectra are compared quantitatively. 

\begin{figure*}
    \includegraphics[width=0.33\textwidth]{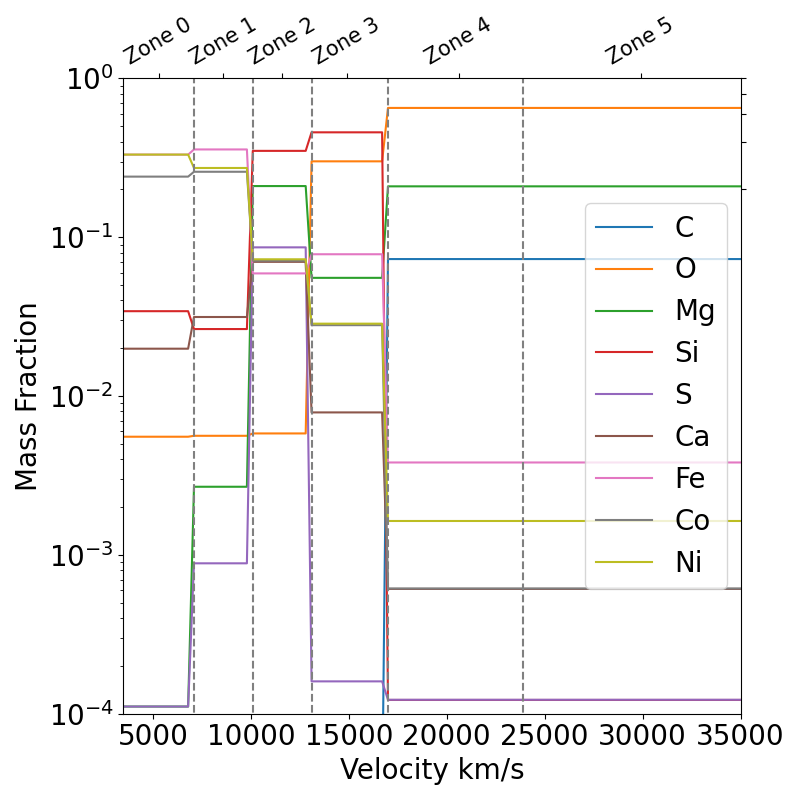}
    \includegraphics[width=0.33\textwidth]{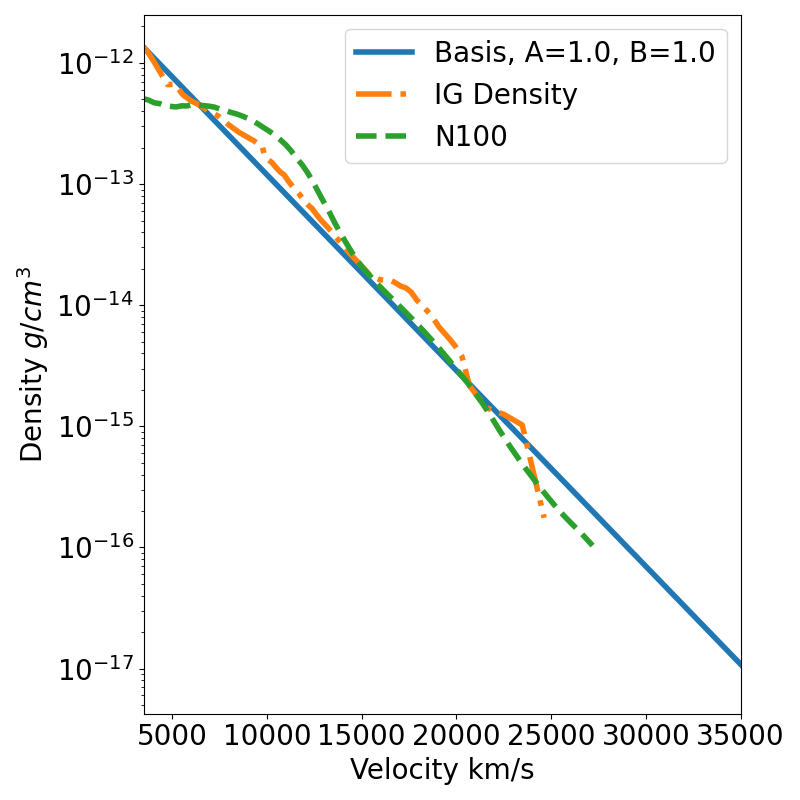}
    \includegraphics[width=0.33\textwidth]{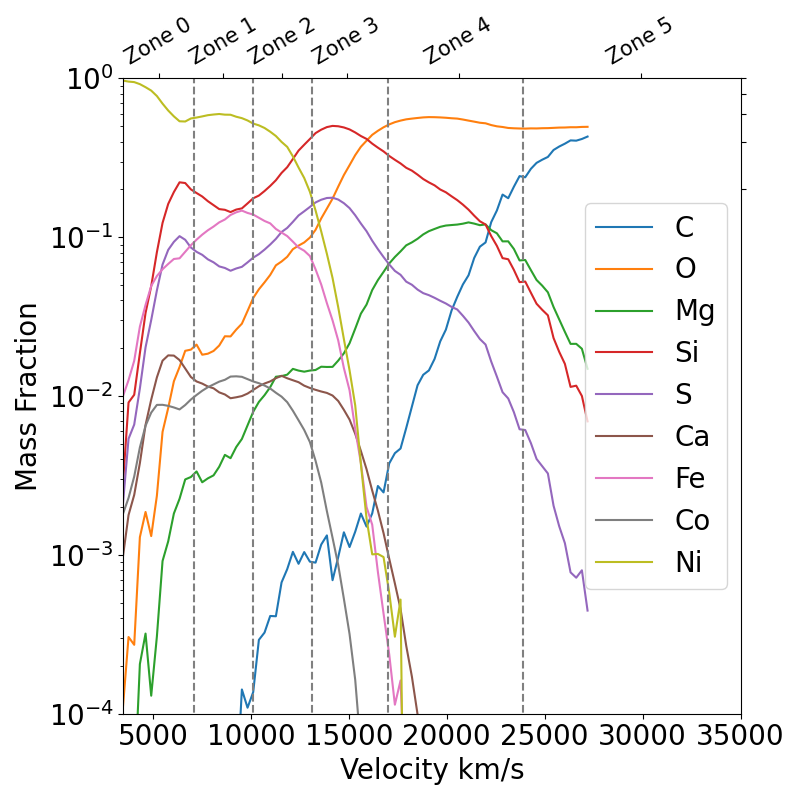}
    \caption{Left: The elemental abundance of the reference model. 
    The label of zones are shown on the top of the panel, and the zone boundaries are marked by the grey dashed lines. 
    Middle: The density structure of the reference model with $A=1$ and $B=1$ in Equation \ref{eq:dens} used in this paper (blue line), the density structure in XLL20 \citep{Xingzhuo2020AIAI} (orange dash-dotted line), and the density structure of the N100 model (green dashed line). 
    The densities shown are at 11.5741 days after the explosion. 
    Right panel: the elemental abundance structure of the N100 model. }\label{fig:IGmodel}
\end{figure*}

\subsection{Simulated Spectral Library}

In this section, we parameterize the supernova ejecta models based on the "IG" model in XLL20 \cite{Xingzhuo2020AIAI}, and randomly sample the parameter space to create a supernova ejecta structure library. 

We treat the SN ejecta density profile as a power-law relation shown below: 

\begin{equation}\label{eq:dens}
    \begin{aligned}
    \rho(v)= & 4.712\times 10^{-14}\left( \frac{t}{11.5741\ days}\right)^{-3} \\ 
             & A\cdot 0.689^{\frac{(v-12500\ km/s) B}{1000\ km/s}}\ g/cm^{3},
    \end{aligned}
\end{equation}
where $A$ and $B$ are dimensionless parameters for the density profile, $v$ is the expansion velocity, and $t$ is the time after explosion. 
When $A=1$ and $B=1$, the density profile is similar to the "IG" model in XLL20 \cite{Xingzhuo2020AIAI} and serves as the reference density profile in this paper. 
A comparison between the density profiles is shown in Figure \ref{fig:IGmodel}. 

We divide the SN ejecta into 6 discrete zones in the velocity space. 
The elemental abundance is assumed to be constant inside a zone. 
There is no layer above 24000 km/s in the XLL20; we extrapolate the high-velocity component from the "IG" model in XLL20 \citep[][]{Xingzhuo2020AIAI}  to serve as the reference elemental abundance in this paper.
The elemental abundance structure and the density profile of the N100 model from \citep{Ropke2012N100} are also shown in Figure \ref{fig:IGmodel} for comparison.

The parameters for generating the spectral library include: 
\begin{itemize}
    \item The density profile parameterized by $A$ and $B$ (see Equation\ref{eq:dens}): $A$ and $B$ are drawn from uniform distributions between  $[0.2,2]$ and $[0,2]$, respectively. 
    \item The abundance of 23 elements in 6 zones of the ejecta structure: The abundance of an element in a given zone is drawn from a Gaussian distribution in the logarithmic space $N(\mu,\sigma^2)$, where $\mu$ is the log of the abundance of the reference elemental abundance structure, and $\sigma$ is set to be 0.5 for all elements. Moreover, the sum of the elemental abundance in each zone is conserved to be 1. 
    \item The time after the explosion: The time after the explosion is drawn from a uniform distribution in $[10,40]$ days.
    \item The supernova luminosity between $6500$ \AA\ and $7500$\AA\ in units of $log_{10}(L_\odot)$ and the location of the inner boundary: To set the parameter ranges of the SN luminosity and the location of the photosphere, we first use the reference ejecta structure (with $A=B=1$ in Equation \ref{eq:dens}) and the 
    extrapolated "IG" abundance structure (from \cite{Xingzhuo2020AIAI}) to calculate a spectral sequence between 10 days and 40 days after the explosion, with varying target luminosities and inner boundary velocities. By comparing the simulated spectra with the observed SNe~Ia spectral energy distribution \citep[e.g.,][]{Hsiao2007Template,HuLSTM2022ApJ...930...70H}, we determine the lower and upper limits of the target luminosities and the locations of the inner boundary at different epochs after the explosion, the results are shown in Table~\ref{table:lumivelo}. 
\end{itemize}


\begin{table}[ht]
    \centering 
    \begin{tabular}{cccc} 
    \hline\hline 
    Time (days) & Luminosity ($log_{10}(L_{\odot})$) & Velocity (km/s) \\ [0.5ex] 
    \hline 
    10 & 8.2-8.35  & 10500-13000  \\
    12 & 8.3-8.4   & 10000-13000  \\
    14 & 8.4-8.5   &  9000-12000  \\
    16 & 8.5-8.6   &  7800-10000  \\
    18 & 8.5-8.7   &  7000-8200   \\
    20 & 8.55-8.7  &  6400-7500   \\
    22 & 8.55-8.65 &  5700-6700   \\
    24 & 8.5-8.65  &  5200-6200   \\
    26 & 8.47-8.62 &  4500-5800   \\
    28 & 8.42-8.6  &  4300-5100   \\
    30 & 8.38-8.52 &  4000-4900   \\
    32 & 8.3-8.5   &  3750-4700   \\
    34 & 8.15-8.45 &  3500-4700   \\
    36 & 8.1-8.35  &  3500-4700   \\
    38 & 8.0-8.3   &  3500-4700   \\
    40 & 7.8-8.2   &  3500-4700   \\[1ex] 
    \hline 
    \end{tabular}
    \caption{The inner boundary velocity and the luminosity range to generate the spectral library. }
    \label{table:lumivelo} 
\end{table}

When generating spectra for the spectral library to be used for neural network training, we sample the day after SN explosion from a uniform distribution in $[10,40]$, then sample a target luminosity in logarithmic space from the uniform distribution between $[L_{min},L_{max}]$, where $L_{min}$ and $L_{max}$ are taken from Table~\ref{table:lumivelo} with linear interpolations for days not included in the Table. 
The inner boundary velocity is also sampled from the uniform distribution in $[V_{min},V_{max}]$, where $V_{min}$ and $V_{max}$ are linearly interpolated from Table~\ref{table:lumivelo}. 

The wavelength range of the model spectra is between $2000$ and $10000\ \AA$ with a total of 2000 wavelength bins. Each wavelength bin is separated by a frequency difference of $5.9958\times 10^{11} \mathrm{Hz}$. A total of 108,389 ejecta models and their corresponding spectra are calculated to form the spectral training set for building the NNs. 

\subsection{Neural Network}

A typical NN consists of an input layer, an output layer, and multiple hidden layers. 
The input layer receives data as input tensors, then the tensors are processed by the hidden layers, and finally output to the output layer. 
Different types of layers, representing different mathematical calculations, can be used in the hidden layers. 
For example, the convolution layer convolves the data tensor with a convolution core, the activation layer applies a non-linear function onto the data tensor, pooling layer reduces the size of the data tensor by binning the adjacent data with the maximum values or the average values, fully-connected layer multiplies a matrix to the input tensor (for a review, see \cite{Lecun2015DeepLearning}). 

The training target of a neural network is to provide predictions ($y_{pred}$) over a variable of our interest that is close to the truth ($y_{true}$), from the given information ($X$). 
Initially, the trainable parameters (i.e. the convolution core in the convolution layers, and the matrix in the fully-connected layers, all denoted as $w$) are randomly assigned. 
The neural network then calculates the output ($y_{pred}$) using the known information $X$ as input, and a loss function (i.e. Mean Squared Error: $Loss=Mean(y_{true}-y_{pred})^2$) is calculated to evaluate the performance of prediction. 
The trainable parameters are updated with respect to the gradient $\partial L/\partial w$ and a pre-defined learning rate, to minimize the loss function. 
For computational efficiency, the dataset is separated into batches, and the trainable parameters are updated at each batch. 
The neural network will browse the data set multiple times, one browse is defined as one epoch. 

Here we have adopted the deep ensemble NNs \citep{Laks2016DeepEnsemble} which has the ability to make predictions and prediction uncertainties simultaneously. 
Comparing to typical NNs such as AlexNet \citep[][]{AlexNet} and VGG16 \citep{VGG16}), the output of deep ensemble NNs is not the just the prediction value, but the prediction mean $\mu$ together with the prediction error $\sigma$, assuming a Gaussian noise distribution. 
The loss function is: 
\begin{equation}\label{eq:loss}
    Loss=Mean\left( ln(\sigma(x))+\frac{(y_{true,i}-\mu(x))^2}{\sigma(x)^2}\right),
\end{equation}
where $x$ is the input 
information of the neural network which includes the spectrum, the time after the explosion, and the density parameters; $\mu$ and $\sigma$ are written in function forms to represent the neural network. 

We use the multi-residual NNs \citep{MResNet,Xingzhuo2020AIAI} for the NN architecture. 
In brief, we introduce a block structure with two convolution layers, a batch normalization layer, and an activation layer, and repeat the block structure 7 times. 
The input of the block structure is the sum of the output of all the previous block structures and the input of the first block structure. 
Moreover, the NN has two inputs, the 2000-element array {\tt\string input\_1} that receives the input spectra with 2000 pixels from 2000 to 10000 $\AA$, and the 3-element array {\tt\string input\_2} that receives the density parameters $A$ and $B$ shown in Equation \ref{eq:dens} and the time after the explosion. 
We use the Scaled Exponential Linear Unit (SELU) function as the activation function for all the hidden layers: 
\begin{equation}
    selu(x)=\begin{cases}
        scale\times x, & (x>0) \\ 
        scale\times\alpha\times (e^{x}-1), & (x\leq 0),\\
    \end{cases}
\end{equation}
where 
$scale=1.05070098$ and $\alpha=1.67326324$. 
The detailed architecture is shown in the \href{https://geronimochen.github.io/images/Model_1_0.h5.svg}{online material}. 

Among all the 108389 spectra, 80\% of them are assigned to the training set, 20\% are included in the testing set. 
The training dataset will be used to update the trainable parameters, while the testing dataset is only used to monitor the loss function after every epoch. 

Note that not all the observed spectral data cover the entire UV-optical-IR wavelength from 2000 to 10000 \AA. 
Data augmentation is applied to the dataset to cope with the problem. 
In this augmentation method, we duplicate both the training and the testing data set by making 5 copies of them. 
The spectra in the first copies of the training and the testing data set remain the same as the spectra in the original spectral library. 
For each spectrum in the other 4 copies, we randomly select an observational facility (telescope and instrument) that was used to acquire the spectral observations of SNe in the WISeREP database, then apply the wavelength coverage of that facility onto the spectrum. 
Furthermore, each spectrum is normalized by dividing the spectrum by the average pixel value, in order to avoid overflow in the NNs. 

We used the {\tt\string adam} algorithm \citep{adam} to update the trainable parameters, with a batch size of 200. 
The training process consists of two phases with different learning rates. 
In the first phase, the learning rate is 0.00003 and decays $10^{-6}$ per batch. 
The loss in Equation \ref{eq:loss} is calculated on the training data set and the testing data set respectively after every epoch. 
When the loss on the testing data set does not decrease in the recent 10 epochs, we cease the first phase and keep the model with the smallest loss for the second phase. 
In the second phase, the learning rate is 0.0000003 and decays $10^{-6}$ per batch. 
We also monitor the loss on the training data set and the testing data set every epoch. 
When the loss on the testing data set does not decrease in the recent 10 epochs, we cease the training and keep the NN model with the smallest loss on the testing data set. 

\begin{figure}
    \plotone{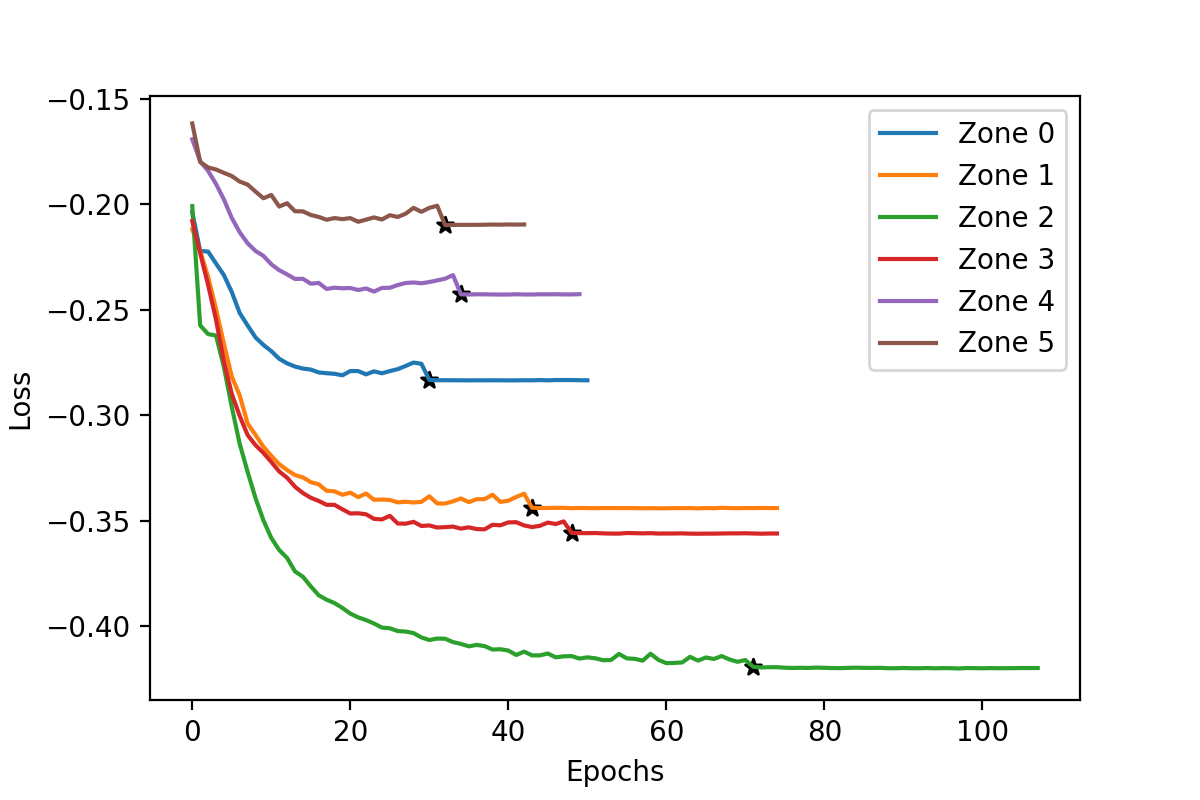}
    \caption{The learning curves of the 6 NNs for elemental abundance predictions in Zones $0 - 5$, showing the losses on the testing data set at different epochs. 
    The black stars show the starting points of the second phase in the training process. 
    The losses for zones 0 to 5 are shown with blue, orange, green, red, purple, and brown lines, respectively. }\label{fig:neuroLoss}
\end{figure}

\begin{figure}
    \plotone{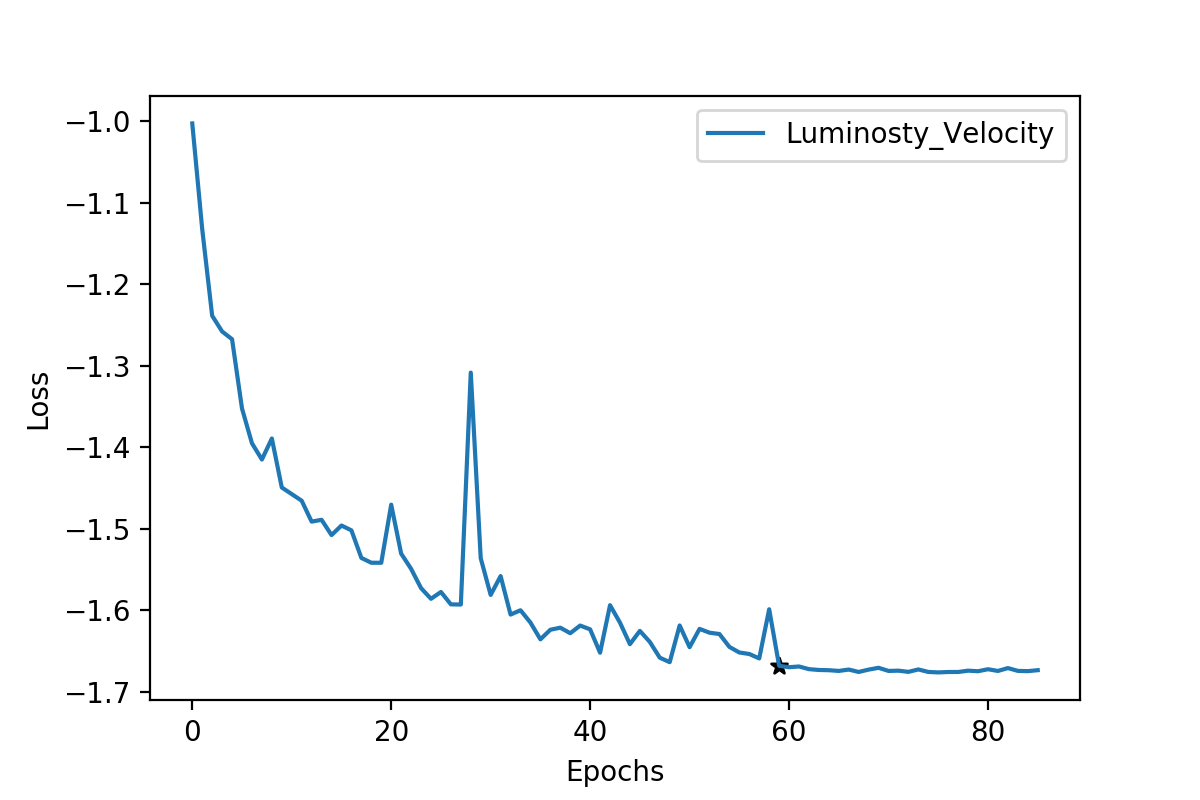}
    \caption{The learning curve of the NN predicting the luminosity and inner boundary velocity. The line shows the losses with the testing dataset at different epochs. 
    The black star shows the beginning of the second phase in the training process. }\label{fig:neuroLossLV}
\end{figure}

We trained 7 sets of NNs in total, the first six NNs  are designed to predict 30 elemental abundances (from H to Zn) in Zones $0 - 5$, and the seventh NN is designed to predict the inner boundary velocity and the luminosity. 
The seventh NN has two outputs, while the other six NNs have 30 outputs. 
The loss on the testing data set is shown in Figure~\ref{fig:neuroLoss} and Figure~\ref{fig:neuroLossLV}. 

\subsection{Density Estimation}

In XLL20 \citep{Xingzhuo2020AIAI}, we applied a grid search method to determine the inner boundary velocity and the target luminosity, while applying a rigid density profile for both the model spectral training set and density profile for applications to the observed spectra. 
With the increased number of SNe to fit, and the two extra density parameters to be determined, the grid search is too computationally expensive for the current research. 
A more efficient search algorithm is designed for this study.
For an observed spectral sequence of an SN~Ia, we first use the NNs to predict the time sequences of the elemental abundances, the luminosities, and the velocities of the inner boundary for a given grid of density parameters and the phase of each spectrum.  
We use TARDIS to re-calculate the simulated spectral sequences for this density parameter grid. 
The simulated spectral sequences are compared with the observed spectral sequence using the mean squared error as the merit function: 
\begin{equation}\label{eq:MSEAB}
    MSE(A,B)=\sum_i\sum_\lambda (F_{i,obs}(\lambda)-F_{i,sim}(\lambda,A,B))^2,
\end{equation}
where $i$ is the index of spectra in the spectral sequence, $F_{i,obs}(\lambda)$ is the $i$-th observed spectrum, $F_{i,sim}(\lambda,A,B)$ is the $i$-th spectra in the simulated spectral sequence using the density parameters $A$ and $B$. 
The simulated spectra are masked according to their wavelength coverage of the observed spectra with the missing wavelength regions set to have zero weights and normalized by dividing the spectra by the average pixel value. 
The observed spectra are re-sampled with the wavelength grid of the simulated spectra, and also normalized by their average pixel values. 

Equation~\ref{eq:MSEAB} has only two free parameters $A$ and $B$ to be determined. The best $A$ and $B$ parameters that minimize Equation~\ref{eq:MSEAB} are calculated through a grid search but still involve a large number of spectral model computations. 
We have applied the following method for further improvement.
The method has 4 iterations, each iteration calculates the spectral sequence in a different sub-grid. 
In the first iteration, the sub-grid is set to be $A=[0.5,1.1,1.7]$, $B=[0.5,1.1,1.7]$. 
In the second iteration, the sub-grid is specified as $A=[a-0.2,a,a+0.2]$, $B=[b-0.2,b,b+0.2]$, where $a$ and $b$ are the parameters set with the smallest mean squared error from the first iteration. 
In the third iteration, the sub-grid is replaced by $A=[a-0.2,a,a+0.2]$, $B=[b-0.2,b,b+0.2]$ with $a$ and $b$ being the values that give the smallest mean squared errors from the second iteration.  
In the fourth iteration, the sub-grid is refined to  $A=[a-0.1,a,a+0.1]$, $B=[b-0.1,b,b+0.1]$, where $a$ and $b$ are the parameters set with the smallest mean squared error from the third iteration. 
Finally, the parameter set with the smallest mean squared error from all the sub-grids will be adopted to be the density parameter of the supernova. 

In this grid search strategy, a typical number of TARDIS models to be calculated is $(9+8+8+8)\times N_{obs}$, where 9, 8, 8, and 8 are the number of grid points in the four iterations described above, and $N_{obs}$ is the number of observed spectra. 
For a well-observed SN with $\sim$ 30 spectra (e.g., SN~2005cf), the total computation time can exceed 1000 CPU hours. 
Therefore, we limit the number of observed spectra used in the grid search to about 10 based on the quality of the spectral data. 
For the well-observed SNe, we only select the spectra from ground-based telescopes with apertures larger than 3 meters in diameter and the Hubble Space Telescope (HST). The spectra are visually inspected to eliminate the data with obvious abnormal flux calibrations. 

\section{Testing}

\subsection{Error Estimate}\label{sec:errest}

\begin{figure}
    \plotone{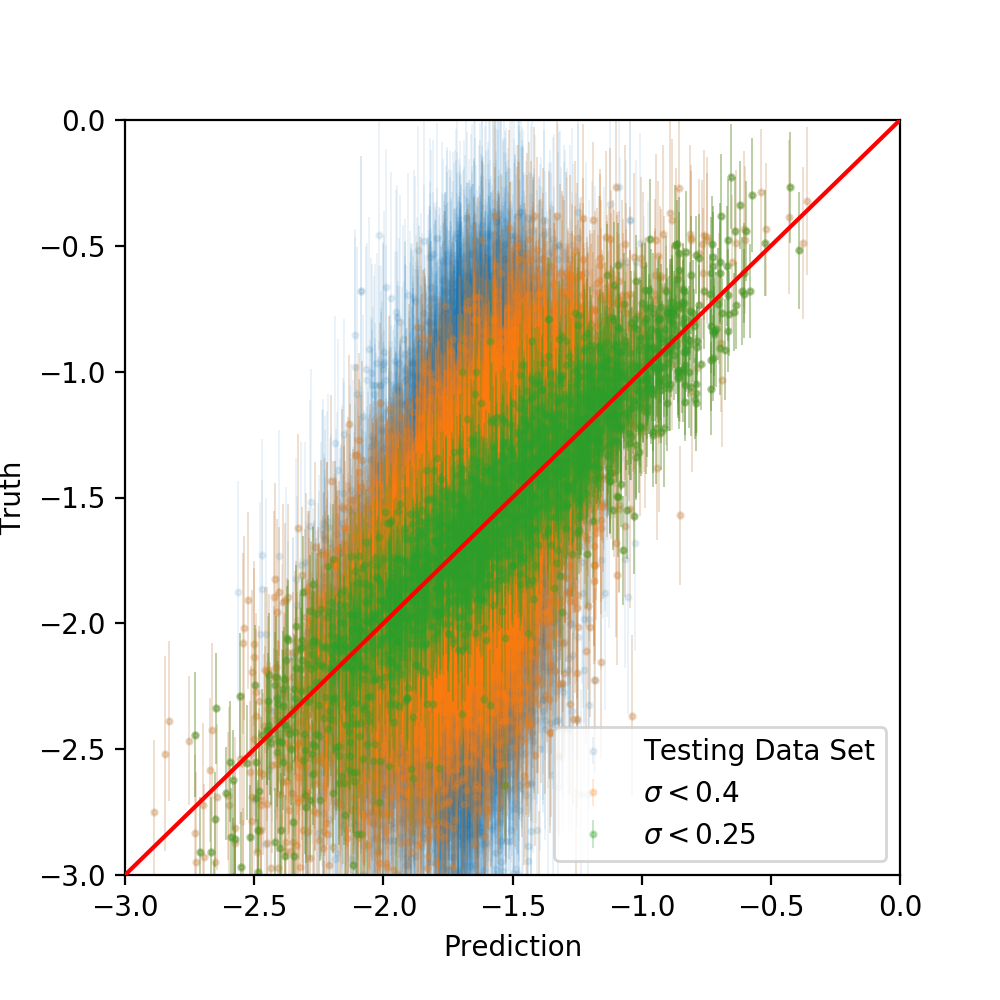}
    \caption{A comparison between the truth values and the predicted values of Ni abundance in Zone 3. 
    The unit of both of the axes is the Ni mass fraction in logarithmic scale ($log_{10}(Ni)$). 
    Data points with large predictive errors ($\sigma \geq 0.4$) are in blue color, data points with medium predictive errors ($0.4 > \sigma \geq 0.25$) are in orange color, data points with small predictive errors ($\sigma \leq 0.25$) are in green color. 
    A reference line (Prediction=Truth) is shown in red color. 
    For illustration purposes, we only include 20000 data points randomly selected from the testing data set. }\label{fig:residual}
\end{figure}

In Figure \ref{fig:residual}, we use the neural network to predict the Ni abundance in Zone 3 using the testing data set, together with its associated predictive error ($\sigma$). 
We notice when the predictive error is small, the predicted value is consistent with the true value. 
With increasing values of the predictive error, the predicted values systematically move away from the true values. 
When the predictive errors are larger than 0.4 dexes, the predicted values become insensitive to the truth values and take values in the range between -2.0 and -1.5. 
Therefore, we use $\sigma$ as a criterion to measure the goodness of prediction on the observed spectra. 

\subsection{Spectral Fitting}

\begin{figure*}
    \includegraphics[width=0.9\textwidth]{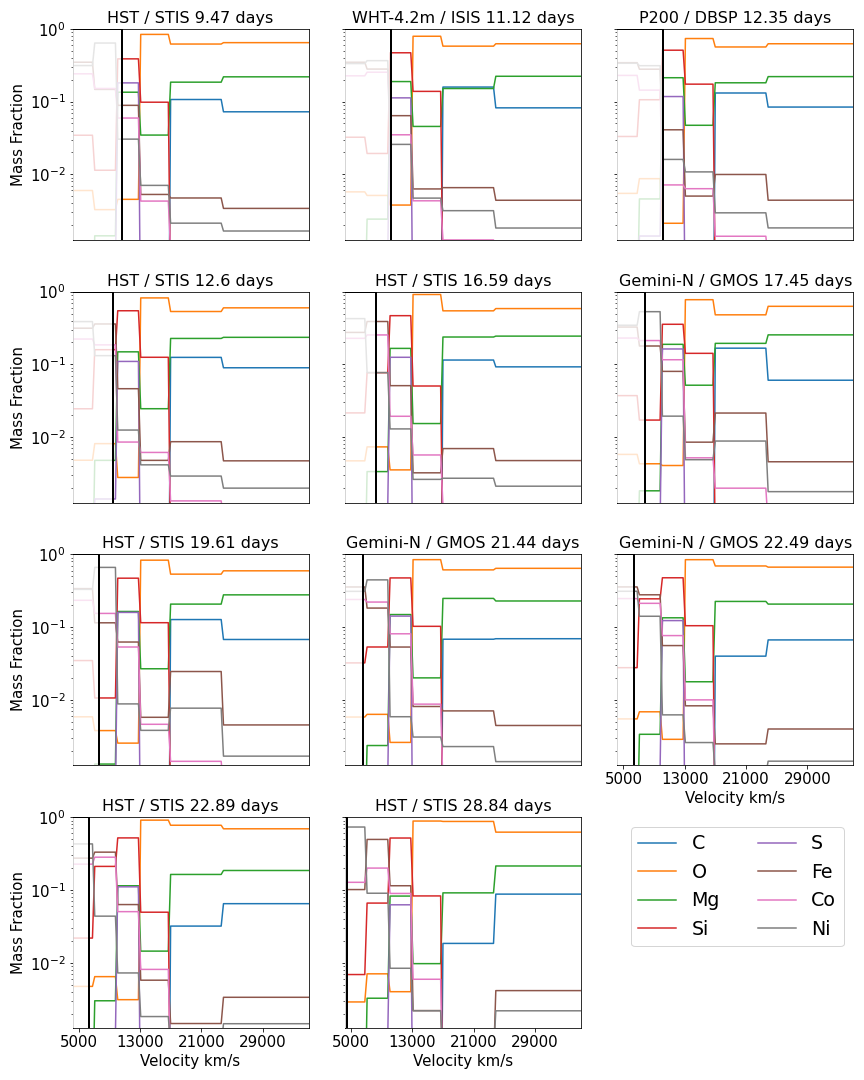}
    \caption{The element abundance time sequence predicted from the SN2011fe spectra with density parameters $A=1.0$ and $B=1.1$. 
    The sources of the data and the time after the explosions are shown as panel titles. 
    The thick vertical black lines mark the velocities of the inner boundary which are also one of the parameters predicted by applying the neural networks to the observed spectra. 
    Elemental abundances below the inner boundaries are insensitive to the observations of their corresponding epochs.  }\label{fig:11feElemSeq}
\end{figure*}

\begin{figure*}
    \centering
    \includegraphics[width=0.8\textwidth]{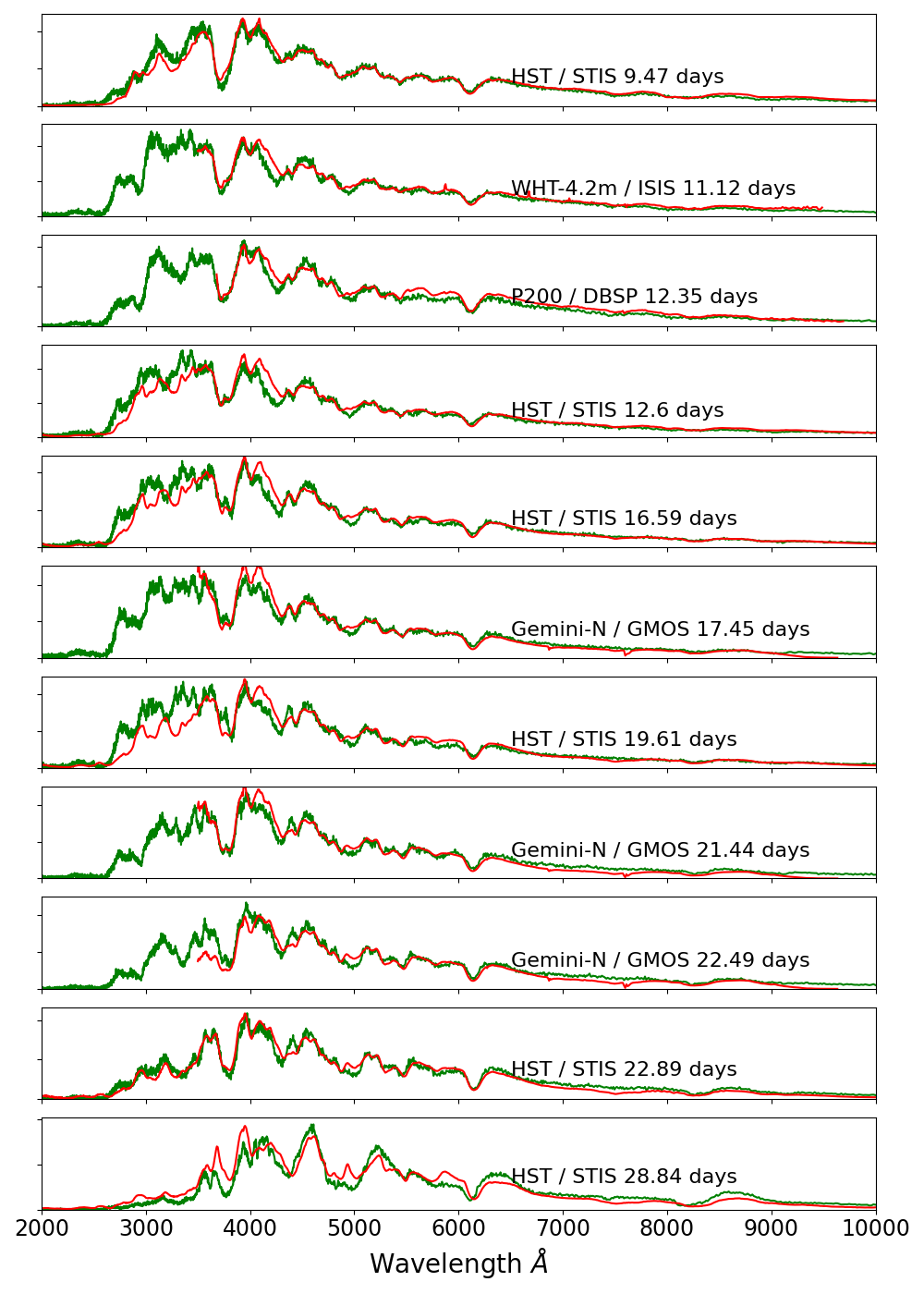}
    \caption{A comparison of the simulated spectra (green lines) using the density parameters $A=1.0$, $B=1.1$, and the observed spectra (red lines) of SN~2011fe. 
    The sources of the data and the epochs after the explosion are shown above the data of each epoch. 
    All the observed spectra are de-redshifted and with Milky Way and host galaxy dust removed as in {\tt\string kaepora} \citep{siebert2019Kaepora}. }\label{fig:11feSpecSeq}
\end{figure*}

The parameters describing the density profile of the ejecta are determined from a grid search for the optimal values using Equation~\ref{eq:MSEAB} for SN~2011fe.  
The optimal values of the parameter $A$ and $B$ are found to be 1.0 and 1.1, respectively. 
Figure~\ref{fig:11feElemSeq} shows the elemental abundances derived from the observed spectral sequence of SN~2011fe. 
Figure~\ref{fig:11feSpecSeq} compares the best-fit spectra from the neural networks and the observed spectra. 

We notice the spectra of SN~2011fe at 12.6 days, 16.59 days, 19.61 days, and 28.84 days show  larger discrepancies in the UV than other epochs. Overall the quality of the spectral fits is worse in the ultraviolet than in the visible wavelength. 
This may be due to several reasons. 
First, the density profile used in this simulation is a simple exponential function which can be inappropriate for fitting the UV spectral region which is extremely sensitive to the ejecta moving at high velocities. 
Secondly, the non-local thermodynamic equilibrium (NLTE) effect becomes more significant at the late time SNe Ia spectra, while TARDIS only uses {\tt\string nebula} approximation to account for this effect. Thirdly, the assumption that all photons initiate from a sharp inner boundary is an over-simplification which can also affect the reliability of the models, especially in the UV.  Nonetheless, the AIAI approach as adopted in this study makes it possible to derive model-based quantities that are sensitive to the chemical structures of the ejecta, although these derived quantities are dependent on the adopted models.


\subsection{Ni Element Sensitivity}

\begin{figure*}
    \minipage{0.5\textwidth}
        \includegraphics[width=\textwidth]{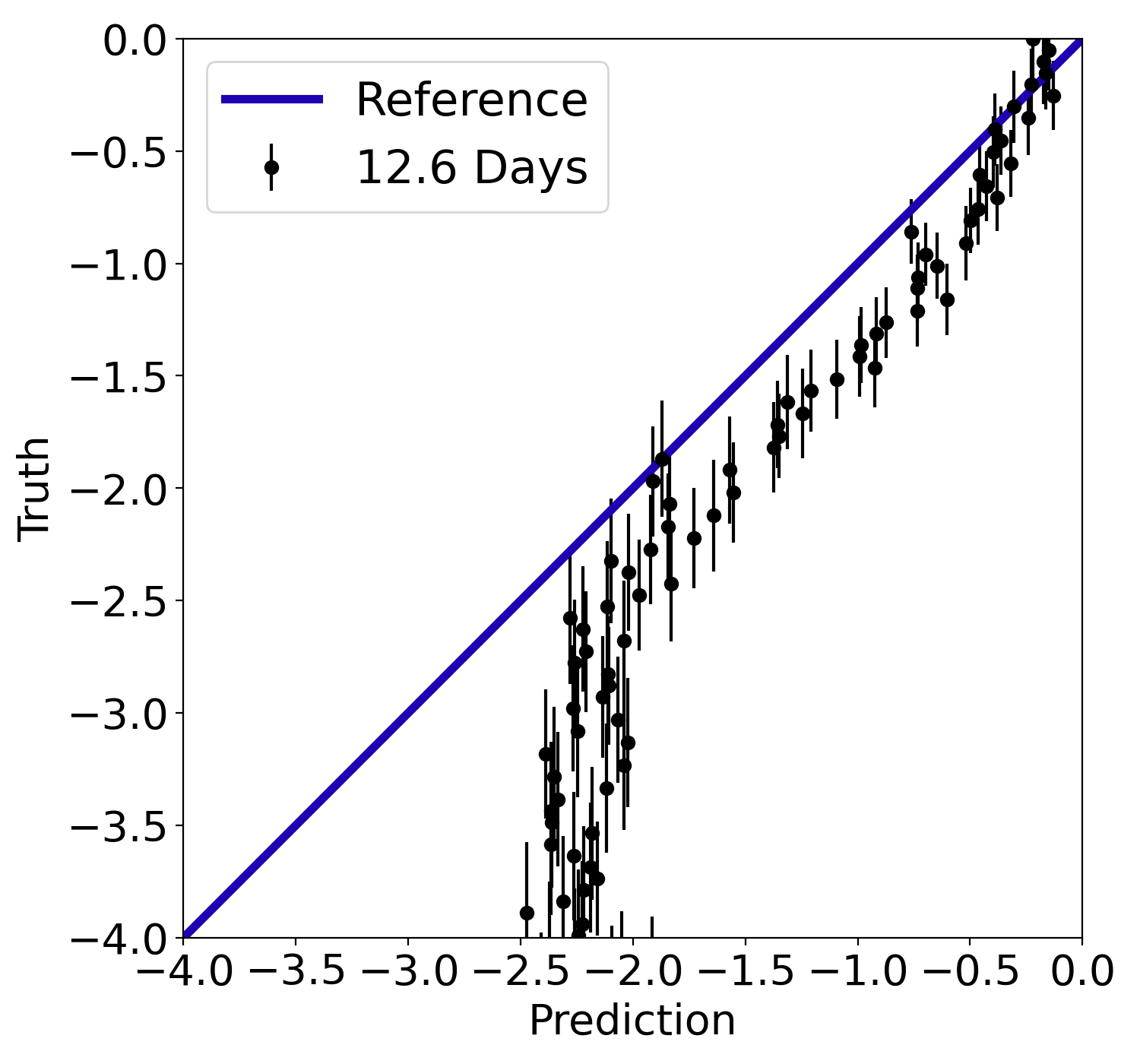}
    \endminipage\hfill
    \minipage{0.5\textwidth}
        \includegraphics[width=\textwidth]{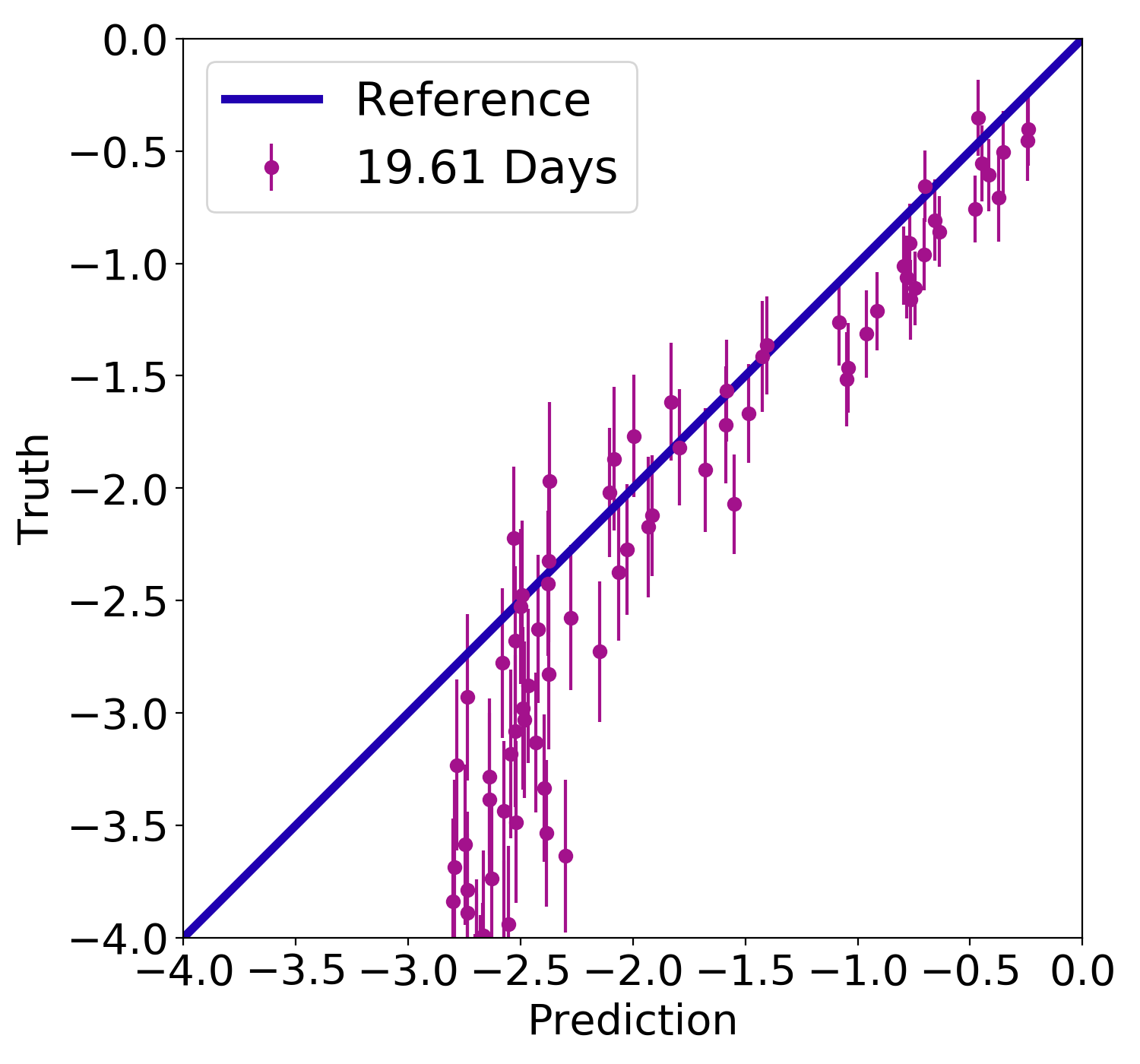}
    \endminipage\hfill
    \minipage{0.5\textwidth}
        \includegraphics[width=\textwidth]{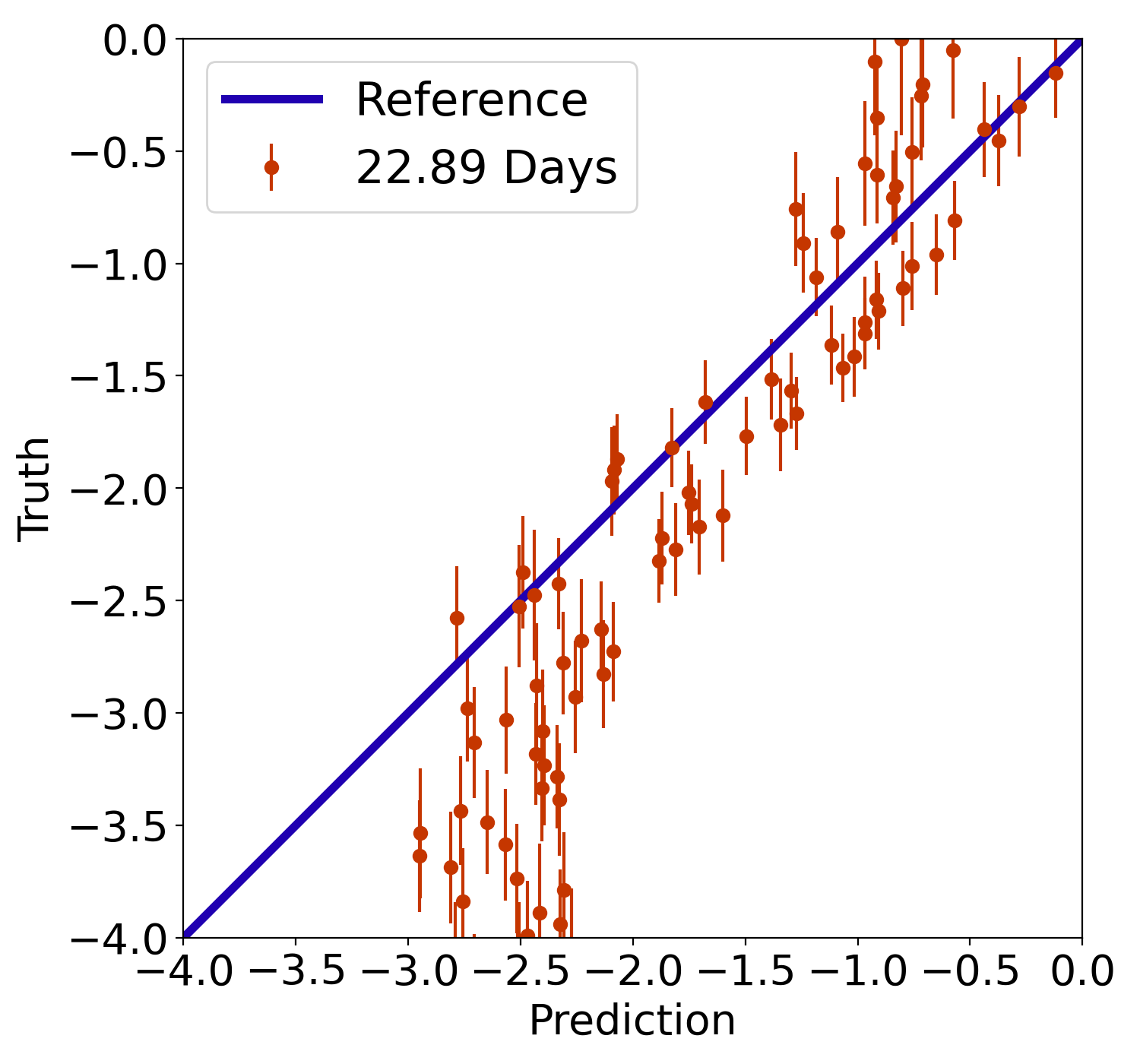}
    \endminipage\hfill
    \minipage{0.5\textwidth}
        \includegraphics[width=\textwidth]{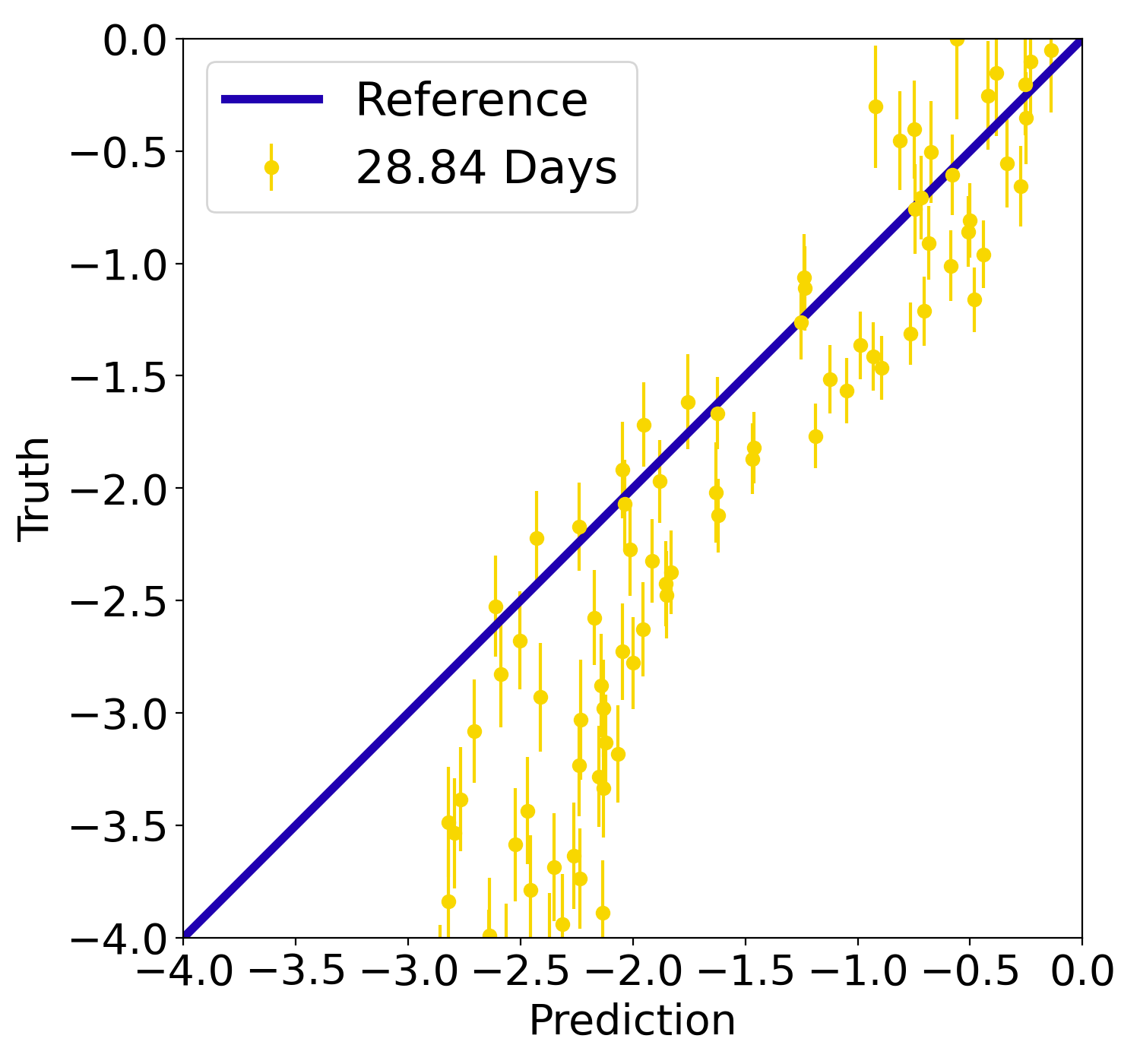}
    \endminipage\hfill
    \caption{The predicted value compared to the truth value of the neural network. The horizontal and vertical axes show the Truths and Predictions, respectively, of the logarithmic Ni mass fraction $log_{10}(Ni)$ in the velocity range $13000 - 17000 km/s$ (Zone 3). 
    The spectra used for these predictions are generated from TARDIS based on the ejecta structure derived for SN~2011fe but varying the Ni mass fraction in the velocity range $13000 - 17000 km/s$. 
    The blue lines show Prediction = Truth. 
    The times after the explosion of the simulated spectra are labeled in the upper left of each figure. 
    }\label{fig:NiSeqSpec}
\end{figure*}

In this section, we use the SN ejecta structure predicted from the HST spectra of SN~2011fe at 12.6 days, 19.61 days, 22.89 days, 28.84 days to simulate 4 spectral sequences with different Ni abundance between $13000-17000\ km/s$ (Zone 3). 
In these models, we allow the mass fraction of Ni in Zone 3 to vary from $10^{-4}$ to $10^{0}$ while keeping the relative ratio of other elements in Zone 3 unchanged. Other model parameters such as the density structures are kept the same for all models. 
With this simulated spectral sequence, we employ the NNs constructed in the above section to derive the predicted elemental abundances. This serves as a test of the capability of the neural network in recovering the assumed model abundances before they can be applied to observational data. 
Figure \ref{fig:NiSeqSpec} shows the predicted and the true Ni abundance in Zone 3 of the spectral sequences. 
The typical prediction errors are $\sim 0.2$ dexes, and the predicted values are in general consistent with the truth when Ni abundance is larger than $10^{-2.5}$ for the spectra in all four phases. 

\section{Results}\label{sec:result}

\begin{figure*}
    \plottwo{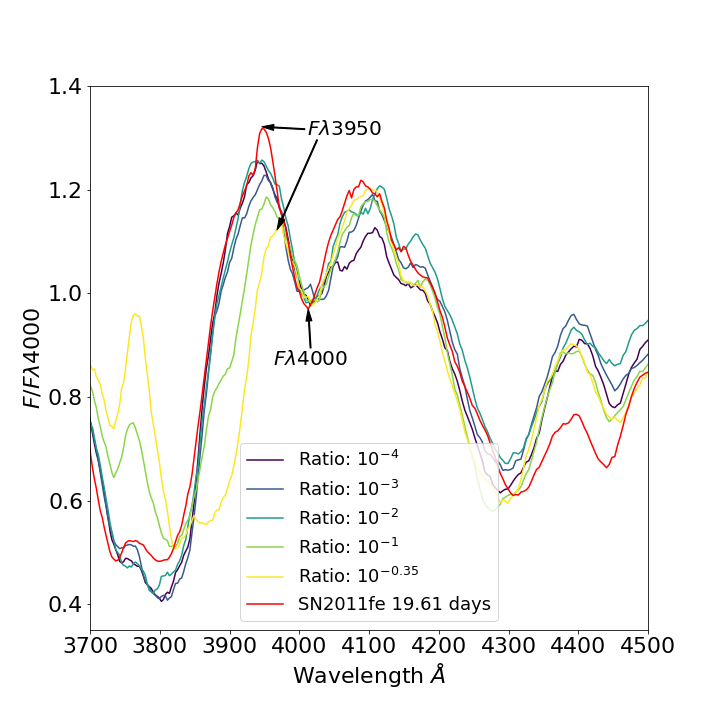}{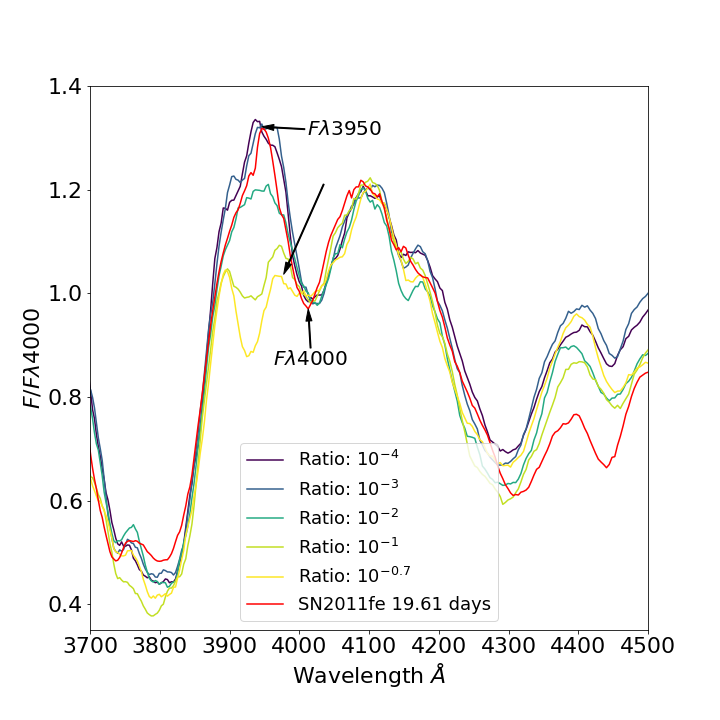}
    \caption{An illustration of measuring the spectral line flux ratio $FRNi$. 
    The red lines are the spectra of SN2011fe at 19.61 days after the explosion. 
    The lines of the different colors are the simulated spectra using TARDIS. 
    The simulated spectra on the left and right are computed with varying  Ni mass fractions in the velocity ranges $13000-17000$ km/s (Zone 3) and  $10000-13000 km/s$ (Zone 2), respectively. 
    The Ni mass fractions are shown in the legends of the two figures. 
    All the spectra are normalized to the local minimum flux near $4000\AA$. 
    The black arrows indicate the local minimum and maximum that are used to measure the spectral line flux ratio $FRNi = F(\lambda 3950)/F(\lambda 4000$). 
    }\label{fig:FRNi}
\end{figure*}

To identify the spectral lines from Ni, we apply our algorithm to the spectra of SN~2011fe at 19.61 days after the explosion  taken by the Hubble Space Telescope (HST) to derive the chemical structure of the ejecta. 
Based on this ejecta structure, we artificially modify the elemental abundance in the velocity range between $13000-17000km/s$ (denoted as Zone 3) and the velocity range $10000-13000km/s$ (denoted as Zone 2) by changing the Ni fraction while re-scaling the abundances of the rest of the elements to keep the density profile conserved. 
Figure~\ref{fig:FRNi} shows the simulated spectra with  Ni abundance in Zone 2 and 3 varying from $10^{-4}$ to $10^{-0.7}$ and $10^{-4}$ to $10^{-0.35}$, respectively.
We notice that Ni has a strong effect on the spectral features at $3750-3950 \AA$ which can be attributed to the blueshifted absorption of the $\mathrm{Ni\ II}\ 4067 \AA$ transition lines. 
This is the only line in the optical in which the strength of Ni II absorption can be directly measured. This line is especially prominent when the Ni fraction is above $10^{-2}$ in these velocity regions. 
Note also that the pseudo equivalent width (pEW) of the Si II absorption line at $\sim 4000 \AA$ is affected by the $\mathrm{Ni\ II}$ transition lines. 
It has been shown in previous studies \citep{Silverman2012BSNIP,Walker2010SNLS} that the pseudo equivalent width (pEW) of the Si II absorption line at $\sim 4000 \AA$ is correlated to the light curve stretch parameter and the absolute magnitude of SNe~Ia. 
It is likely that the measurement of the pEW of the 4000 \AA\ feature is capturing the combined effect of the $\mathrm{Ni\ II}$ and the $\mathrm{Si\ II}$lines in this wavelength region. 

Figure~\ref{fig:NiSeqSpec} shows the neural network predictions on the Ni abundance in Zone 3 based on a set of simulated spectra using the ejecta structure for the models of SN~2011fe, but with varying the Ni abundances in Zone 3. 
It can be seen that the predictions on the Ni abundance in Zone 3 are consistent with the truth values at different times after explosion for abundance values larger than $\sim 10^{-2.5}$. The linear relation is lost when the abundance drops below $\sim 10^{-2.5}$, indicating the sensitivity limit of the models. Figure~\ref{fig:NiSeqSpec} also shows an offset between the prediction and truth for day 12.6 which becomes less prominent at later epochs.  
We surmise the low Ni abundances are unable to produce spectral features stronger than the Monte-Carlo noise in the simulations, and that there are some hidden degeneracies in the model that are still not fully understood. A likely reason is that the electron scattering at zone 3 is still important at Zone 3 at the early epochs which reduces the sensitivity to Ni abundances in the spectral formation.


We focus on the flux ratio between the emission feature around $3950 \AA$ and the absorption feature around $4000\AA$, because the two spectral features can be identified in most of the SNe Ia optical spectra, and the wavelength range is small such that the flux ratio is less affected by the dust extinction effect. 
Firstly, we measure the flux of the local minimum close to $4000\AA$ and define the minimal flux as $F\lambda 4000$. 
Then, we measure the nearest local maximum flux to the blue-side of the local minimum and define the local maximum flux as $F\lambda 3950$. 
Finally, we define the flux ratio $FRNi=F\lambda 3950 / F\lambda 4000$, and use this $FRNi$ as a spectral indicator that can be related to the Ni abundances. 

We searched all the available SNe~Ia spectra from WISeREP \citep{wiserep} and kaepora \citep{siebert2019Kaepora} and found 616 spectra from 153 SNe~Ia, which were observed between -10 days and 20 days relative to the B-band maximum time with the rest frame wavelength covering $3800-7000 \AA$, for our analysis. 
We also collected the $\Delta m_{15}$ parameters of these SNe from kaepora \cite{siebert2019Kaepora} or calculated the $\Delta m_{15}$ parameter using the light curves from kaepora \cite{siebert2019Kaepora} or Open Supernova Catalog \footnote{https://sne.space/}. 
All the spectra are converted to rest frame and corrected for  dust extinction effects using the data from kaepora \cite{siebert2019Kaepora}. 
The times after the explosion of these SNe are calculated assuming the $B$-band maximum time is 19 days after the explosion. 
Moreover, because the inner boundary velocity of TARDIS falls in Zone 2 for most of the SN models at the epochs between $-10$ and $0$ days relative to their B-band maximum, we will only use the Ni abundance in Zone 3 which is less affected by the location of the TARDIS inner boundary for the early-phase spectral models as measures of Ni abundances when comparing models with observations. 

\begin{figure*}
    \minipage{0.33\textwidth}
        \includegraphics[width=\textwidth]{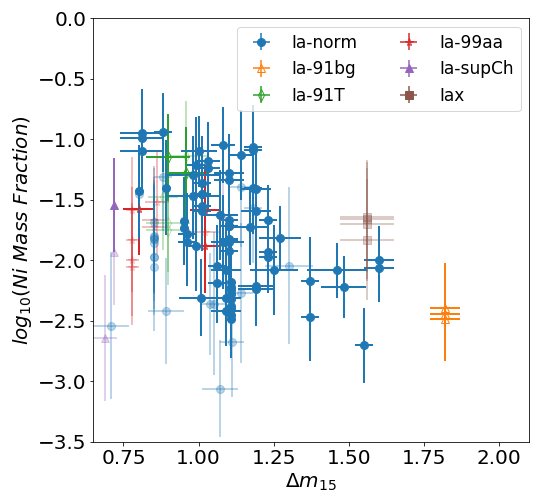}
    \endminipage\hfill
    \minipage{0.33\textwidth}
        \includegraphics[width=\textwidth]{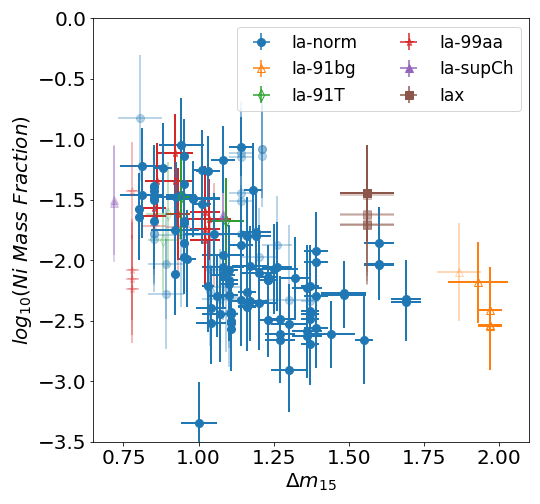}
    \endminipage\hfill
    \minipage{0.33\textwidth}
        \includegraphics[width=\textwidth]{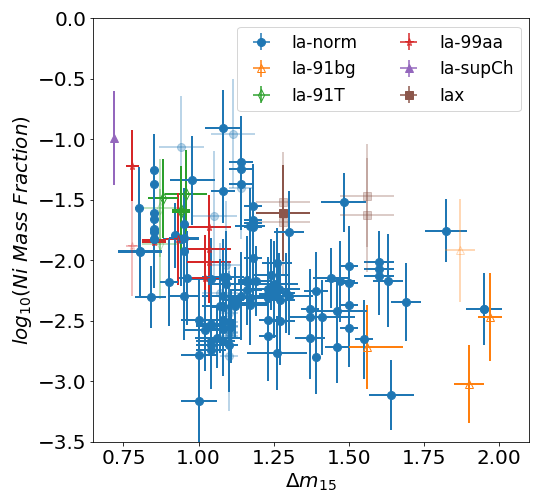}
    \endminipage\hfill
    \caption{The relation between $\Delta m_{15}$ and the Ni abundance in Zone 3 at different time bins. 
    Left panel: between 9 and 12 days from the time of the explosion. 
    Middle panel: between 12 and 15 days from the time of the explosion.
    Right panel: between 15 and 18 days from the time of the explosion. 
    $y$-axis is the Ni mass fraction in Zone 3 in the logarithmic scale. 
    Spectroscopically normal SNe~Ia are marked with blue solid circles, Ia-91bg-like SNe are marked with orange hollow triangles, Ia-91T-like SNe are marked with green diamonds, Ia-99aa-like SNe are marked with red solid stars, Ia-super-Chandrasekar SNe are marked with purple solid triangles, and Iax SNe are marked with brown squares. 
    Data points with large errors on the $y$-axis ($\sigma >0.4$) are transparent. }\label{fig:dm15Ni3}
\end{figure*}

In Figure \ref{fig:dm15Ni3}, we show the relation between $\Delta m_{15}$ and the Ni abundance in Zone 3 of the selected SNe. 
The Figure shows clear correlations between $\Delta m_{15}$ and the derived Ni mass fraction in Zone 3 for the spectra between 9 and 12 days and between 12 and 15 days after the explosion. 
The correlation is much weaker in the data between 15 and 18 days after the explosion, as the Ni abundances are significantly lower than in earlier epochs and for most of the SNe they are close to the detection limit. 
The Ia-99aa subtype SNe and the Ia-91T subtype SNe are found in the regions with larger Ni mass fraction and smaller $\Delta m_{15}$. 
The Ia-91bg subtype SNe shows large $\Delta m_{15}$ and low Ni mass fraction, consistent with them being underluminous due to the deficiency of $^{56}Ni$ in the ejecta as the source of energy. 
The Iax subtype SNe are comparatively isolated from the rest of the SNe Ia, showing large $\Delta m_{15}$ and large Ni mass fraction. 
Moreover, the Ni abundance in Zone 3 does not show a significant decrease for type Iax SNe. If confirmed by observations and models, this indicates that the Ni in Type Iax may be primarily stable isotopes of Ni. 

\begin{figure*}[htb!]
    \minipage{0.33\textwidth}
        \includegraphics[width=\textwidth]{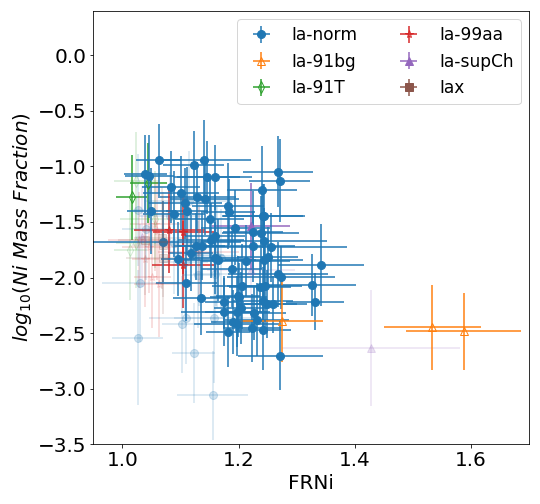}
    \endminipage\hfill
    \minipage{0.33\textwidth}
        \includegraphics[width=\textwidth]{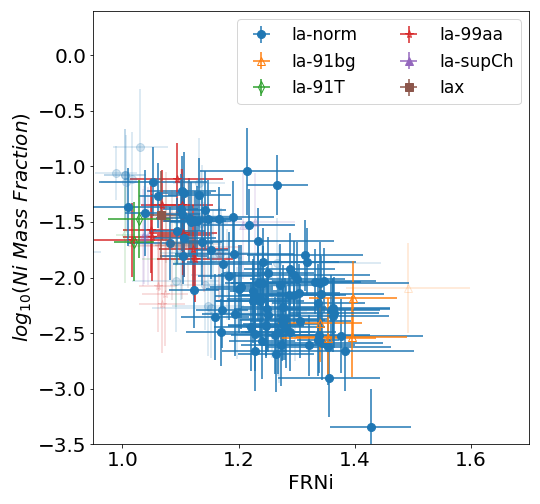}
    \endminipage\hfill
    \minipage{0.33\textwidth}
        \includegraphics[width=\textwidth]{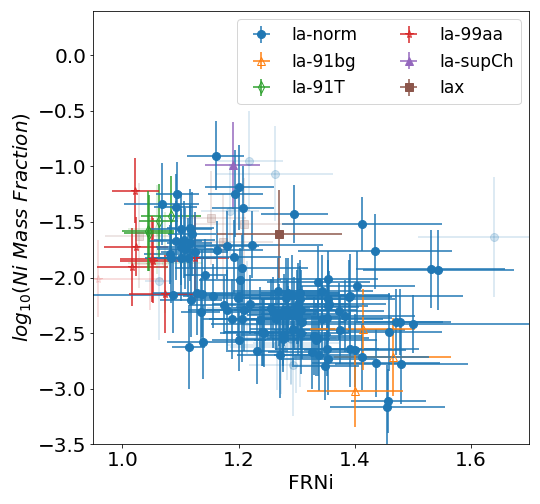}
    \endminipage\hfill
    \caption{The relation between the spectral line ratio ($FRNi$) measured from the observed spectra of SNe and the Ni abundance in Zone 3 determined by applying the neural network models to the observed data. The three panels are for observations in different time windows after the explosion. 
    Left panel: between 9 and 12 days after the explosion. 
    Middle panel: between 12 and 15 days after the explosion.
    Right panel: between 15 and 18 days after the explosion. 
    The vertical axes show the Ni mass fraction in Zone 3 on a logarithmic scale. 
    Different subtypes of SNe Ia are labeled with different colors and marks, the color and mark encodings are the same as Figure \ref{fig:dm15Ni3}. 
    Data points with large errors on the vertical axis ($\sigma >0.4$) are transparent. 
    This figure includes the following sub-Types of SNe~Ia: normal (blue), SN~1991bg-like (orange), SN~1991T-like (green), SN~1999aa-like (red), super-Chandrasekhar (purple), and Iax (brown).}\label{fig:FRNiNi3}
\end{figure*}

The relation between the Ni abundances in Zone 3 and the spectral line ratio $FRNi$ is shown in Figure \ref{fig:FRNiNi3}. 
We notice $FRNi$ shows a linear correlation to the logarithmic Ni abundance in Zone 3 $log_{10}(Ni)$, especially for the spectra between 12 and 15 days. 
Such a linear correlation is weaker for the spectra between the phase 15 and 18 days, probably because  the Ni abundances have decreased below the detection limit for most of the SNe. 
The Ia-99aa and Ia-91T subtype SNe show higher Ni abundances and lower $FRNi$ parameters compared to spectroscopically normal SNe Ia. 
The Ia-91bg subtype SNe show lower Ni abundances and large $FRNi$ parameters. 

\begin{figure}[htb!]
    \plotone{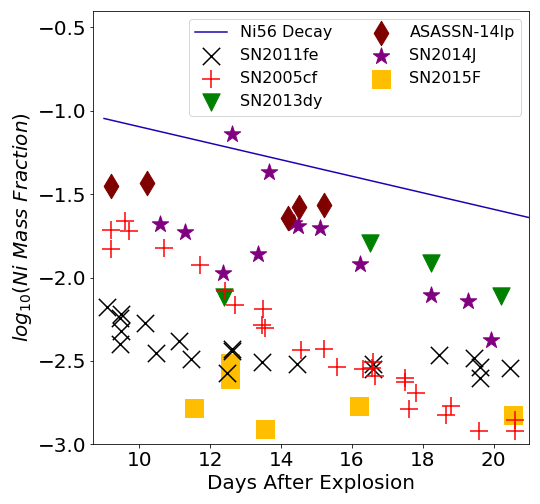}
    \caption{The Ni abundance evolution with time after SN explosion for different SNe~Ia. 
    Y axis is the Ni mass fraction in Zone 3 on a logarithmic scale. 
    The ultramarine line is the theoretical $^{56}Ni$ abundance changing with time. 
    Six well-observed type Ia-normal SNe: SN2011fe (black cross), SN2005cf (red plus), SN2013dy (green triangle), ASASSN-14lp (brown diamond), SN2014J (purple star), SN2015F (yellow square) are shown with different markers. 
    }\label{fig:NiCurve}
\end{figure}

In Figure \ref{fig:NiCurve}, we show the time evolution of the Ni abundance in Zone 3 for SN~2011fe, SN~2005cf, SN~2013dy, ASASSN-14lp, SN~2014J, and SN~2015F. 
The Ni abundances in SN~2011fe and SN~2015F are close to the detection limit, and only SN~2011fe shows the signal of Ni at an early phase. 
SN~2005cf shows a distinct decay of Ni abundance in Zone 3 and the decay rate is slightly larger than the $^{56}Ni$ decay rate. 
The overluminous SN~2013dy and ASASSN-14lp exhibit high Ni abundance in Zone 3, while the Ni abundance decay is not obvious due to inadequate data quality and poor sampling time. 
SN~2014J shows a Ni abundance as high as SN~2013dy or ASASSN-14lp, and the Ni abundance decay trend is observed between $15-20$ days after the explosion, but the data before day 14 show a large scatter. 

\section{Concolusion}\label{sec:conclusion}

We employ the program TARDIS to build a set of NNs to calculate the Ni element in SN Ia ejecta. 
We measure the Ni abundances and their time evolution in the 13000 - 17000 km/s region of 153 SNe~Ia, and correlate the Ni abundance to the spectral feature around 3950 $\AA$ and the supernova light curve parameter $\Delta m_{15}$. 
The neural network prediction on the Ni in the 13000 - 17000 km/s region can reach 0.2 dex accuracy on the simulated SN~2011fe spectra, and the sensitivity can reach $10^{-2.5}$ of Ni mass fraction. 
On the observed spectra, we found the predictive error of most of the predictions is smaller than 0.4 dex, and the predicted Ni abundance is loosely correlated to the nickel spectral line feature $FRNi$ and the light curve parameter $\Delta m_{15}$, especially in the early phase spectra, probably due to more $^{56}Ni$ exists in the early phase SNe Ia ejecta. 
The Ni spectral line features $FRNi$, and the ejecta structures predicted by the neural network could be further used in cosmological studies to estimate the $H_0$ value more accurately. 

We also notice the simulated spectra based on the neural network predictions do not perfectly fit the observations, especially in the ultraviolet wavelength. 
Moreover, the Ni abundance evolution of several well-observed SNe only loosely observes the theoretical $^{56}Ni$ decay curve. 
We surmise these defects could due to the crude approximation of NLTE calculation in the radiative transfer simulation program TARDIS, and the over-simplification of the theoretical SN ejecta density profile. 
The deviation of TARDIS model from Sedona and CMFGEN is found to be the largest in the UV wavelength range \citep[see Fig. 12 in,][]{Blondin2022arXiv220911671B}. 
The TARDIS code generates photons from an inner boundary, and the other part of the ejecta only reprocess photons from the inner boundary. 
Sedona made the improvement of more realistically generating photons in the Monte-Carlo simulation, but Sedona models are much more costly to compute. 

With AI, it is possible to 'calibrate' codes with simplified approximations to make them agree with results from codes with more complete treatments of physical processes. The physical model-based parametrization of the observational data of SNe~Ia as developed in this study can be extended to include other radiative transfer codes \citep[e.g.,][]{Hoeflich1996Hydra,Hillier1998CMFGEN,Baron1998Phoenix,KasenSedona2006ApJ...651..366K}. Such parametrization is different from what has been done so far based on empirical models of SN~Ia luminosity-distance relations \citep[e.g.,][]{phillips1993dm15relation,Tripp1998A&A...331..815T,JhaMLCS2007ApJ...659..122J,GuySALT,CMAGICIWang_2003,Wang_CMAGICII2006}. For future projects of supernova cosmology, the theoretical model-based parametrization has a different sensitivity to the systematic errors of supernova evolution, it may prove to be important for controlling the systematic evolutions of the age and metallicity of SN~Ia progenitors, especially when very high redshift ($z\sim 3-6$) SNe~Ia are involved \citep{LuJia2022arXiv221000746L}.

It is also important to have a fast code to generate surrogate models of 3-D radiative transfer to account for the asymmetric geometry of the ejecta as found from spectropolarimetry observations \citep{Wang:1996ApJ...467..435W,Wang2008ARA&A..46..433W,Cikota:2019MNRAS.490..578C,Yang2022arXiv220812862Y} and theoretical models \citep[e.g.,][]{NeopaneFisher2022ApJ...925...92N}. 
These works will be reported in a forthcoming study (Chen et al. in prep.). 


\clearpage

\begin{acknowledgements}
    Portions of this research were conducted with the advanced computing resources provided by Texas A\&M High-Performance Research Computing. 
    This research used resources of the National Energy Research Scientific Computing Center (NERSC), a U.S. Department of Energy Office of Science User Facility operated under Contract No. DE-AC02-05CH11231. 
    LW acknowledges the NSF grant AST-1817099 for support on this work.
    This paper made use of {\tt\string netron}, an open-source program for neural network visualization \href{https://github.com/lutzroeder/netron}{https://github.com/lutzroeder/netron}. 
    This research used TARDIS, a community-developed software package for spectral synthesis in supernovae \cite{tardis, tardisSoftware}. 
    The development of \textsc{Tardis} received support from the Google Summer of Code initiative and from ESA's Summer of Code in the Space program. \textsc{Tardis} makes extensive use of Astropy and PyNE. 
\end{acknowledgements}



\end{document}